**Spiders and Crawlers and Bots, Oh My: The Economic Efficiency and Public Policy of Online Contracts that Restrict Data Collection**

<u>I. BACKGROUND</u>

Recent trends reveal the search by companies for a legal hook to prevent the undesired and unauthorized copying of information posted on websites. This quest has sparked a fundamental controversy over the ownership of information and the Internet itself. Some believe that information published on websites should be used only in a manner sanctioned by the publisher, so that "businesses can proceed without fear of unwanted trespassers who will steal or profit from the fruits of [the business'] labor."[1] Many argue, however, that the efficient exchange of factual information, unhampered by any legal or technological barrier, unquestionably benefits society and weighs strongly against the enforceability of any restrictive mechanism.[2]

Especially troublesome to those who struggle against this unwanted copying of website information are software robots, small programs that automatically and rapidly search, copy and retrieve information from websites.[3] Prevalent and controversial among software robots are shopbots or pricebots, which comb through commercial websites, extracting pricing and product information.[4]  Typically, shopbots are used in conjunction with a metasite, a website that displays prices from a variety of vendors for an identical item.[5]  While shopbots and metasites tend to revolve around pricing information, these systems have been employed to compile and exploit information ranging from the schedules of ticketed events to the contact information of website proprietors.[6]

---

[1] *See* May Wong, Associated Press, *eBay, Bidder's Edge End Lawsuit* (March 2, 2001), *available at* http://dailynews.yahoo.com/h/ap/20010302/tc/ebay_settlement_1.html (quoting Jay Monahan, eBay's legal counsel for intellectual property)

[2] *See* Brief of Amici Curiae Mark A. Lemley et al. at 3-8, Ebay, Inc. v. Bidder's Edge, Inc., 100 F. Supp. 2d 1058 (N.D. Cal. 2000) (No. C-99-21200RMW).

[3] *See* Stephen T. Middlebrook & John Muller, *Thoughts on Bots: The Emerging Law of Electronic Agents,* 56 BUS. LAW. 341, 342-43 (2000). A robot is a program that navigates the World Wide Web's hypertext structure, recursively retrieving documents. Robots are also referred to as "bots," "spiders," and "web-crawlers." Search engines illustrate a fundamental use of robots; these robots traverse the Internet, recursively following hyperlink after hyperlink and indexing the web pages at the end of the links. The indexed web pages are then stored in the search engine's database so that a user may later query the search engine to discover the indexed page. While robots are not intrinsically harmful to the internet, as in a properly functioning search engine robot, a poorly implemented robot or a flood of identical robots can tax the Internet's infrastructure. *See* Martijn Koster, *A Standard for Robot Exclusion, at* *http://info.webcrawler.com/mak/projects/robots/norobots.html*. Different than robots, but also relevant to the struggle against illegitimate copying are unauthorized deep links. Hyperlinks are software instructions sent to a networked computer to retrieve data. With websites, hyperlinks are the text and pictures, which when clicked, take the clicker to a different web-page. Deep links take the user to an interior page of another's website. Therefore, if a website wishes for all its users to enter through a homepage, an unauthorized deep link may thwart this policy. *See* Michael F. Finn and James M. Assey, *Handling Unauthorized Deep Links: Legal Issues and Practical Strategies,* 4 ACCA DOCKET 36, 38 (2000).

Websites attempting to thwart price/information indexers may invoke technological methods to prevent access by robots.[7]  First, a website may incorporate a robot exclusion header, a text file that indicates that the site does not allow unauthorized robotic activity.[8]  However, compliance with the Robot Exclusion Standard is entirely voluntary; a robot must be programmed to read the header and conform to its control directives before searching a website.[9]  Second, if a website detects a robot's presence from repeated and rapid requests generated from a single server, the website may then block inquiries from that server's Internet Protocol address.[10]  However, attempts to block queries from a specific IP address often prove unsuccessful, because robot information requests may be effected through proxy servers, which frustrate efforts to locate the originating IP address.[11]  Finally, a website might employ password technology to limit its contents to authorized viewers.[12]  However, password mechanisms currently fail to distinguish between human users and robots, and furthermore, this procedure may hinder legitimate access.[13]  No technological method, therefore, currently exists to effectively prevent robot searches of websites while maintaining the site for sanctioned public use.  Furthermore, the risk with any technological advancement remains, that the blocked robot will find a way to circumvent the new technological barrier, causing the searched site to become even more restrictive, in a "technological arms race."[14]

---

[4] *Frictions in Cyberspace*, THE ECONOMIST , Nov. 20, 1999, *available at* http://www.economist.com/displayStory.cfm?Story_ID=346410; *see also infra* Part II.C.

[5] *See Frictions in Cyberspace supra* note 4.  Metasites offer little original content, but rather aggregate and organize the content of other websites.  *See infra* Part II.C.

[6] *See generally* Ticketmaster Corp., v. Tickets.com, No. 99CV7654, 2000 WL 1887522 (C.D. Cal. Aug. 10, 2000) (presenting a claim against website that placed deep hyperlinks to plaintiff's ticketing and event information web-pages); Register.com v. Verio, Inc., No. 00 CIV. 5747, 2000 WL 1855145 (S.D.N.Y. Dec. 8, 2000) (presenting a claim against website that used data robots to extract names, physical addresses, email addresses, and telephone numbers for patrons of the plaintiff).

[7] *See* Maureen A. O'Rourke, *Shaping Competition on the Internet: Who Owns Product and Pricing Information,* 53 VAND. L. REV. 1965, 1984-85 (2000).

[8] *See* Martijn Koster, *A Standard for Robot Exclusion, at* http://info.webcrawler.com/mak/projects/robots/norobots.html.  To implement a robot exclusion header, the server must create a text file available on the "[local URL]/robots.txt."  The file consists of only two elements: 1) the types of robots that the exclusion header is aimed at, and 2) the portion of a URL that is not to be visited by the robot.  This approach has generated momentum because of its ease of implementation and the ability of a robot to access the information with only one query.  *See id.*

[9] *See id.*

[10] *See* Ebay, Inc. v. Bidder's Edge, Inc., 100 F. Supp. 2d 1058, 1061 (N.D. Cal. 2000); *See* David Kramer and Jay Monahan, *Panel Discussion: To Bot or Not to Bot: The Implications of Spidering*, 22 HASTINGS COMM. & ENT. L.J. 241, 245-46 (2000).  Every server maintains a unique Internet Protocol ("IP") address.  When a server queries another server, the requesting computer must furnish its IP address so that the information requested may be sent.  If a server detects quickly repeated queries from a specific IP address, this may indicate that the server is hosting a robot.  Once robotic activity is confirmed, the requested computer may simply block queries from that specific IP address.

In response to the failure of these technological barriers, websites have asserted a variety of legal claims to protect their posted information.[15]  In particular, recent cases illustrate the use of the common law doctrine of trespass to chattels to block further searches by data robots.  Trespass to chattels occurs when one interferes with the possession of personal property of another, and thereby proximately causes injury.[16]  In *Ebay, Inc. v. Bidder's Edge, Inc.* and *Register.com v. Verio, Inc.*, two different courts found that the use of shopbots to scour websites constituted a trespass to chattels when the unauthorized robotic activity drained the plaintiffs' system resources, thereby causing injury.[17] In a criticized aspect of these two cases, the courts disregarded that Bidder's Edge's and Verio's shopbots generated a maximum of 1.53% and 2.3% of the system queries, respectively, and essentially aggregated the hypothetical drain on system resources if multiple robots combed the plaintiffs' websites simultaneously.[18] In this fashion, the courts satisfied the requirement of actual injury with the possibility of future harm.

---

[11] *See Ebay* 100 F. Supp. 2d at 1061; Kramer & Monahan, *supra* note 10, at 245-46.  Proxy servers act as a locus for outgoing server queries and preserve system resources by centralizing outgoing and incoming data.  Generally, use of proxy servers is limited to local users.  Often due to inadequate maintenance, however, remote users may employ proxy servers.  By exploiting this feature, remote users can route outgoing queries through a proxy server, thereby effectively concealing their location.  A server hosting a robot may send its robotic queries through a proxy server and attach the proxy server's IP address to all of its inquiries.  Any server detecting robotic activity will be lead to the proxy server and not to the actual server hosting the robot.

[12] *See* Kramer & Monahan, *supra* note 10, at 245-46.

[13] *See id* at 245-46*; see also* O'Rourke, *supra* note 7, at 1985.

[14] *See* Kramer & Monahan, *supra* note 10, at 247; *see also* O'Rourke, *supra* note 7, at 1985.  As those who try to thwart technological restriction mechanisms become more resourceful, so will websites then add further layers of technological protection to block the newly learned techniques.  The addition of layer upon layer of technological protection influences and limits innocent website use.  *See* O'Rourke, *supra* note 7, at 1985.  Currently, companies exist that create and sell software with the specific purpose of evading any kind of defensive mechanism implemented to block data robots.  *See* Kramer & Monahan, *supra* note 10, at 247.

[15] Several of these legal claims have generally proved to be unreliable and are beyond the scope of this paper.  First, while copyright law may provide a remedy for some robotic replication of website data, most shopbots only extract unprotected factual information.  *See* Feist Publications, Inc. v. Rural Telephone Service Co., 499 U.S. 340, *passim* (1991).  Second, while the Digital Millenium Copyright Act criminalizes "technological measure[s] that effectively control access to a work protected by copyright," it is unclear that circumventing either a Robot Exclusion Header or IP address blocking will invoke the DMCA's protection.  17 U.S.C. § 1201(a)(1)(2000).  Finally, while the Computer Fraud and Abuse Act criminalizes the transmission of a program with reckless disregard of a substantial and unjustifiable risk to the recipient computer, those who attempt to invoke the CFAA may find it difficult to demonstrate damages (as *Ebay* illustrates) and the intent to defraud.  18 U.S.C. § 1030(a)(5)(B) (2000).

Given the precarious injury criterion of the trespass to chattels doctrine, dicta in both *Ebay* and *Register.com* reveal that contract law may provide a less demanding legal method of preventing the search of websites by data robots. In *Ebay*, the court notes that Ebay requires consumers to accept a user agreement, which among other things, prohibits the use of any robot to monitor or duplicate content contained within the website.[19] The court then suggests that while Bidder's Edge never agreed to comply with the user agreement, if it had consented, the terms of the contract would have been binding.[20] In *Register.com,* the court explicitly relies upon the enforceability of a user contract to find that Verio breached the agreement when it employed the robotically extracted information in an unauthorized manner.[21] Furthermore, the court indicated that while Register.com's user agreement fails to prohibit the use of robots to search its website,[22] a contract limiting the use of robots would be enforceable.[23] Therefore, both *Ebay* and

---

[16] Thrifty-Tel v. Bezenek, 54 Cal.Rptr.2d 468 (1996); *see* W. Page Keeton ET. AL., Prosser and Keeton on the Law of Torts, 85-87 (5th ed. 1984).

[17] *See Register.com*, 2000 WL 1855145, at *11; *Ebay*100 F. Supp. 2d at 1070.

[18] *See* Brief of Amici Curiae Mark A. Lemley et al. at 14, Ebay, Inc. v. Bidder's Edge, Inc., 100 F. Supp. 2d 1058 (2000) (No. C-99-21200RMW). The courts found that if the defendant's use of robots were to proceed unchecked, such use would encourage others to engage in similar robot activity. Instead of only one robot indexing Ebay's data, the court hypothesized there may soon be several. And once others begin to use similar robots to collect data from a website, system use will skyrocket, and the website will suffer "irreparable harm from reduced system performance, system unavailability, or data losses." *Ebay* 110 F. Supp. 2d at 1066.

[19] *See Ebay* 100 F. Supp. 2d at 1060.

[20] *See id.* at 1060.

[21] *See Register.com*, 2000 WL 1855145, at *5-8.

[22] Rather, Register.com's agreement forbids using its data to send mass unsolicited commercial advertising. Therefore, it is not the robot gathering of data that Register.com proscribes, but rather the use of the data once it has been collected. It is this aspect of the contract that Verio breached. *See id*. at *5-8.

[23] *See id.* at *8.

*Register.com* signify that online contracts offer a legally viable mechanism to prohibit shopbots from scouring websites and extracting information, regardless of how the information is ultimately used.

If blocking shopbots is as simple as posting a mandatory user agreement on a website, the question arises whether this end result is acceptable and desirable. Part II of this paper argues that contrary to popular belief, enforceability of contracts that restrict shopbots ("robot restriction contracts") will aid in achieving economic efficiency, rather than hindering its progress. Part III of this paper reviews the application of common law contract doctrine and the Uniform Computer Information Transactions Act ("UCITA") to robot restriction contracts. Both the trends in contract common law and UCITA sanction the enforceability of robot restriction agreements. Part IV of this paper integrates the results of the economic efficiency analysis of Part II and the contract law analysis of part III to gauge whether public policy is served by the enforcement of robot restriction contracts. This investigation indicates that the creation of a technical standard to accommodate fair use robotic activity is necessary to preserve public-interest pursuits.

## II. ROBOT RESTRICTION AND ECONOMIC EFFICIENCY

Conventional "brick and mortar" market systems have failed to achieve economic efficiency and perfect competition.[24] However, features of electronic markets and the Internet may reduce the frictions of traditional commerce, which have heretofore prevented an efficient market. Specifically, metasites potentially offer the consumer a costless mechanism to determine pricing and product information for any commodity offered for sale by any vendor.[25] While many have championed metasites as the panacea for imperfect competition, economic modeling and empirical evidence suggest that a rash transition to consumer reliance on metasites may impede economic efficiency, instead of promoting it.[26] Rather, an unhurried transition to a market controlled by metasites will enhance economic efficiency.[27]  In other words, online retailers limiting access to their pricing and product information will further the transition to greater economic efficiency, rather than hinder it.

---

[24] *See generally infra* Part II.A (discussing the elements of economic efficiency).

[25] *See generally infra* Part II.C (discussing reduced search costs of metasites)

[26] Economic modeling illustrates that an abrupt transition to metasites may result in either monopolistic pricing through vendor collusion or cyclical price wars effected through harmful mass-robot proliferation. *See generally infra* Part II.C.1. Empirical evidence suggests that factors unrelated to search costs

## A. Economic Efficiency and Perfect Competition

Richard Lipsey defines perfect competition as "a market structure in which all firms in an industry are price takers and in which there is freedom of entry into and exit from the industry."[28]  Economic efficiency incorporates three elements.[29]  First, *efficiency in distribution* requires that the market distribute products among consumers so that the items are allocated in a manner in which no consumer would prefer a different available item.[30]  Second, *efficiency in production* involves the maximization of the production of desired goods to the extent that resource variables permit.[31]  Third, *consumer sovereignty* requires the production of items most desired by consumers.[32]  A market economy characterized by homogenous products, perfectly informed consumers and the absence of search or transaction costs will generate market prices equal to the marginal cost of production, thereby satisfying the three elements of economic efficiency and creating an environment of perfect competition.[33]

To illustrate why prices equaling the marginal cost of production are necessary for an efficient economy, consider two consumers, *X* and *Y* and two products, *a* and *b*. Furthermore, assume *X*'s marginal rate to substitute *a* for *b* is 2 and *Y*'s marginal rate to

---

contribute to price dispersion.  Thus, eliminating search costs will not produce economic efficiency.  *See generally infra* Part II.C.2.

[27] A controlled transition to metasites will eliminate the dangers of mass-robot proliferation while providing an opportunity for price-dispersing factors unrelated to search costs to dissipate.  Only when these factors fade, will reduced search costs lead to economic efficiency.  *See infra generally* Part II.D.

[28] RICHARD G. LIPSEY ET AL., ECONOMICS 915 (8th ed. 1987); *see infra* note 33 (describing why economic efficiency and perfect competition exist in the same context).

[29] *See* ROBERT DORFMAN, PRICES AND MARKETS 114 (1967).

[30] *See id.* at 115-16.

[31] *See id.* at 116-20.

[32] *See id.* at 120-26.

[33] Karen Clay et al., *Pricing Strategies on the Web: Evidence from the Online Book Industry* *6 (2000) (unpublished manuscript) (*citing* J. Bertrand, *Theorie Mathematique de la Richesse Sociale*, 67 JOURNAL DES SAVANTS 499-508 (1883)).  In the Bertrand economics model, firms produce identical products, consumers are perfectly informed, firms set prices and will elect to set prices at the marginal cost of production.  In the Bertrand model, if a firm charges a price higher than the marginal cost of production it will face zero demand.  Conversely, if a firm charges a price less than the cost of production, it will capture the whole market but will be unable to sustain itself as it generates losses.  Therefore, all firms will fall into equilibria, charging marginal cost of production for all items.  *See id*.

substitute *a* for *b* is 4.  X will then substitute 1*b* for 2*a* with indifference.  However, X would advantageously seek to exchange 1*b* for 3*a*.  Consequentially, *Y* will gain if *Y* can substitute 3*a* for 1*b,* because Y's marginal rate of substitution of *a* for *b* is 4.  This situation, where a mutually advantageous opportunity for substitution exists, is not economically efficient.[34]  Rather, efficient economics requires that all consumers maintain the same marginal rate of substitution for all products.[35]  When all consumers purchase products at the same price, the consumers' marginal rates of substitution necessarily converge.[36]  Thus, a consumer enjoys the greatest utility when the marginal rate of substitution between a pair of products equals the ratio of the products' respective prices.[37]

Given that comprehensive price equality is a necessary condition for economic efficiency, why must that price equal the marginal cost of production for an item?[38] Economic efficiency requires efficiency in production as well as distribution.[39]  As noted above, efficiency in production exists when the production system of all commodities is maximized; in other words, production is efficient when as much of product *a* is created as possible, without lessening the production of product *b*.[40]  Now, assume that 2

---

In a perfectly competitive environment, no vendor has any control over price.  Such a vendor may sell as much or little of an item as it chooses at that item's established price, but may not charge more or less.  Where prices equal the marginal cost of production a vendor has no choice but to sell its goods at that price.  If economic efficiency requires that goods be priced at their marginal cost of production, then economic efficiency necessarily produces perfect competition because vendors must then charge marginal cost of production or fail.  PAUL A. SAMUELSON, ECONOMICS 457 (11th ed. 1980)

Note that economic efficiency and perfect competition will not necessarily be the most profitable framework for a seller.  Rather a perfect monopoly and perfect price-discrimination yield the most profitable result, but both of these characteristics are inherently inconsistent with economic efficiency and perfect competition, because here, sellers would have absolute control over prices.  *See infra,* Part II.C.2.

[34] *See generally* DORFMAN, *supra* note 29, at 115-16 (1967); SAMUELSON, *supra* note 33, at 416-19.

[35] *See* DORFMAN, *supra* note 29, at 115.

[36] What seems like an impossibility, having equal marginal rates of substitution among consumers, becomes a reality with uniform pricing of items.  Consumers select quantities of different items so that their marginal rate of substitution between each pair of items equals the ratio of their prices, thus equating marginal rates of substitution between consumers.  *See id.*

[37] *See id.* at 115.  Thus efficient distribution exists when all consumers share the same marginal rate of substitution between all items, because then no exchange could benefit any consumer without harming others.  *Id.*

[38] Marginal cost is "the increment of Total Cost that comes from producing an increment of one unit of *q*." It can be calculated by subtracting the total costs of adjacent outputs.  SAMUELSON, *supra* note 33, at 442.

[39] *See* DORFMAN, *supra* note 29, at 116.

[40] *See id.* at 116.  Efficient production requires that the value of every product be the same in every firm that employs it, "for then no reallocation of resources among firms could increase the total value of the economy's output or could increase the output of any commodity without reducing the output of some

consumers exist (*X & Y*), who are interested in purchasing *a*, which has a marginal cost of production of 10. *X* is willing to pay up to 10 for *a*; *Y* is willing to pay up to 20 for *a*. If the manufacturer of *a* prices it at any value above 10, then to maximize profits, the manufacturer will produce only 1*a,* where production efficiency necessitates the maximum possible output, or 2*a.* Thus, economic efficiency requires that a product *a* be priced uniformly and at the marginal cost of production.

<div align="center">B. Why Sellers Object to Price Indexing</div>

The question arises why companies would object to shopbot price indexing of their products. One might initially presume that vendors would appreciate the free publicity, which metasites offer. For example, a vendor would benefit when a consumer who had not considered purchasing from the vendor, elects to do so based on metasite use. Numerous factors, however, counteract any shopbot benefit.

First, if contrary to the economic modeling,[41] shopbot and metasite use will inevitably lead to marginal cost of production prices through lower search costs, this in turn will cause reduced seller profits. By preserving elevated search costs, vendors maintain market power, through which they will exploit price dispersion.[42] Specifically, sellers have an incentive to impede the flow of pricing, product, and vendor information when the seller occupies a dominant market position.[43] What would Amazon.com, which charges prices substantially higher than competitors, gain from permitting consumers to view an item by item, price by price and vendor service by vendor service comparison?[44] Amazon.com would, in fact, gain little if it were revealed that it offers the same package of product and services for a higher price. Conversely, reputed brand retailers may fear that metasites will deliver to consumers an inaccurate comparison of sellers.[45] For

---

other." *Id at 144; See also supra,* note 36 (explaining how uniform pricing equates marginal rates of substitution).

[41] *See infra*, Part II.C.

[42] *See* O'Rourke, *supra* note 7, at 1978.

[43] *See id.* at 1978; *see also* J. Bradford DeLong & Michael Froomkin, *Speculative Microeconomics for Tomorrow's Economy*, *in* INTERNET PUBLISHING AND BEYOND 6, 25 (Brian Kahin & Hal R. Varian eds., 2000) (couching the reluctance of established vendors to participate in price indexing as the vendors' exploitation of ignorance about the availability of lower search costs).

[44] Assuming for example that a competitor offers the identical item at a lower price with the same quality of service and warranty.

[45] *See* O'Rourke, *supra* note 7, at 1978.

example, if a metasite only displays two sellers offering the same item for the same price, the seller that has expended significant funds developing an exceptional customer service and delivery system, wastes this cost.[46]

Second, shopbot and metasite use may upset a seller's revenue model.[47] In addition to sales of products, online retailers earn money through the sale of advertising space, as well as commissions received when they link customers to other related specialized websites.[48] Use of metasites necessarily diverts consumer traffic away from the retailers' sites, disrupting the integrity of their revenue models. Metasite use would thus effectively redistribute income from the seller to the middleman, thereby causing increased product prices to compensate for the lack of indirect revenue.

Finally, sellers may fear, as expressed in the trespass to chattels cases, that unauthorized robot use will drain their systems' resources.[49] While current shopbot use, as represented in *Ebay v. Bidder's Edge,* constitutes only inconsequential system use, the proliferation of metasite use suggests that increased robotic activity may tax sellers' systems, causing unnecessary expense.[50]

Therefore, despite the positive aspects of shopbot use for established sellers, several additional factors indicate metasite prevalence will produce negative consequences on their business.

<u>C. Metasite Inefficiency</u>

Given that products priced at their marginal cost of production is a condition for economic efficiency, the question arises whether the reduced search costs associated with shopbots and metasites will inevitably lead to lower prices, unvarying among vendors. Shopbots automatically request pricing and product information from multiple online retailers. In response to a consumer's query of a metasite, a shopbot can within seconds retrieve the product and pricing information for the item selected by the consumer.[51] Thus, a shopbot performs in a matter of seconds what a human consumer would be unable to accomplish in hours. Shopbots, therefore, appear to promise a substantial reduction in the costs of obtaining and distributing product and pricing information, a generally accepted economic boon, reducing market friction and enhancing economic efficiency.[52] While negligible search costs, generating a free flow of product and pricing information, are a prerequisite for prices set at the marginal cost of production, these frictionless searches do not by their nature alone generate economic efficiency. Rather, economic modeling and empirical evidence suggest that: 1) unrestrained shopbot implementation may lead to either monopolistic practices or price wars, which unduly tax the Internet infrastructure, and 2) factors unrelated to search costs preserve price dispersion, frustrating any move toward economic efficiency.

---

[46] This is the weaker of the two arguments. As metasites progress they will offer more and more seller information. Currently, metasites, in addition to price, offer product information, and seller delivery track record, seller warranty and return policy, and customer satisfaction. *See* http://www.MySimon.Com

[47] *See* O'Rourke, *supra*, note 7, at 1982.

[48] *See* DeLong & Froomkin, *supra* note 43, at 29-32; O'Rourke, *supra* note 7, at 1982.

[49] *See Ebay 100* F. Supp. 2d at 1071.

[50] *See id.* at 1071; *Supra* Part II.B.1

    1.  Economic Modeling of Unmonitored Shopbot Proliferation

Traditional economic models commonly assume that consumer search costs are negligible.[53] Furthermore, in a market system of homogeneous products and inconsequential search costs, the conventional Bertrand economics model expects that price competition among vendors in an efficient market will reduce prices to the marginal cost of production, with sellers generating zero profits.[54] While such a result is commonly considered desirable,[55] positive search costs have until now resulted in above marginal cost pricing. It has therefore been theorized that the radical decline in search costs, resulting from shopbots and metasites, will lead to the destabilization of oligopolistic pricing practices and eliminate seller profit.[56]  However, economic modeling suggests that a mass destabilization will bring adverse economic consequences, resulting in higher prices through formation of new oligopolies or through a mass encumbrance of the Internet's infrastructure.

---

[51] Websites like www.shopper.com and www.mysimon.com collect, sort, and display pricing, product, and other information on items ranging from groceries to consumer electronics.[51] WWW.Shopper.com claims to compare 1,000,000 prices of 100,000 different computer components. For example when I entered on March 26, 2001 that I was searching for a VHS version of *Mary Poppins* (one of my favorite films), Mysimon.com presented me with a list of 34 results. The list included the name of the vendor, a merchant review, and the price.

[52] *See* DeLong & Froomkin, supra note 43, at 18-35.

[53] *See* SAMUELSON, *supra* note 33, at 39.

[54] *See supra* note 33; *id* at 39; Michael D. Smith, et al., *Understanding Digital Markets: Review and Assessment*, *in* UNDERSTANDING THE DIGITAL ECONOMY, *4 (Erik Brynjolfsson & Brian Kahin eds. 1999), *available at* http://ecommerce.mit.edu/papers/ude.

[55] *See* generally JACK HIRSCHLEIFER, PRICE THEORY AND APPLICATIONS 234 (4th ed. 1988); SAMUELSON, *supra* note 33, at 456-74 (discussing how imperfect competition harms consumerism and resource allocation); Yannis J. Bakos, *Reducing Buyer Search Costs: Implications for Electronic Marketplaces*, 43 MGMT. SCI. 1676, *passim* (1997).

To illustrate why unrestrained shopbot mechanics may result in either monopolistic pricing practices or substantial burdens on the Internet, consider a market with a product $a$, offered for sale by several vendors, $S$, to a significantly larger number of potential purchasers, $B$.[57]  The value of $a$ to a particular purchaser $b$ is $v_b$.  Therefore, a consumer purchases $a$ when the price is less than or equal to $v_b$, thus obtaining a utility of $v_b$-$p$, when a transaction occurs and a utility of zero otherwise.[58]  Further, assume that consumers contemplate the price of a product based on 1 of 2 different approaches: 1) the consumer will purchase from the first seller that offers $a$ for a price less than or equal to $v_b$, or 2) the consumer will search all vendors and select the vendor that has priced $a$ at the lowest value, assuming the price is less than or equal to $v_b$.[59]  Note that option 2 is available only with an Internet shopbot/metasite scheme and that in a traditional market, consumers would be forced to select option 1 or a drastically diluted form of option 2, thus fundamentally altering the model.  A particular seller's profit is equal to the difference in price and cost multiplied by the demand for $a$ of that seller.[60]  The demand of a particular seller becomes a probability function of two components: a) consumers of type 1 have a $1/S$ probability of selecting the specific seller $s$, and b) consumers of type 2 will select a seller $s$, based on the probability that $s$ offers $a$ at a lower price than other sellers,[61] or, if priced at the same low value as other sellers (including at the marginal cost of production), then 1/(all sellers who have achieved the lower price).[62]

---

[56] *See id.* (asserting that the rise of the electronic marketplace will fundamentally change the three primary functions of the marketplace: a) to match buyers and sellers, b) to facilitate the exchange of information, goods, and services, and c) to provide an institutional infrastructure); Robert Kuttner, *The Net: A Market Too Perfect for Profits*, BUSINESS WEEK, May 11, 1998, at 20.

[57] Jeffrey O. Kephart & Amy R. Greenwald, *Shopbot Economics,* *4, *available at* http://www.research.ibm.com/infoecon/paps/html/aa99_shopbot/aa99_shopbot.html (2000).  Kephart and Greenwald provide a thorough economic modeling of electronic marketplaces and shopbots.

[58] *See id.* at *4.

[59] *See id.* at *4.  The buyer population is composed of both type 1 and type 2 consumers with type one consumers initially outnumbering type 2 consumers.  The distribution of type 1 and type 2 consumers may be derived in multiple ways but both types will always initially exist.

[60] *See id.* at *4.

When all purchasers are of type 1, seller collusion effects an oligopolistic pricing scheme above the marginal cost of production.[63] When all purchasers are of type 2, we arrive at the result of the traditional economic model, where all consumers are perfectly informed of all vendors, resulting in *a* priced at the marginal cost of production.[64] However, the shopbot/metasite scheme promises a balance between type 1 and type 2 consumers, as a shift to shopbot use will occur gradually rather than suddenly and completely. What then is the effect of this mix of consumers? Computations based on the above model reveal that as expected, vendor competition will lead to decreased prices through shopbot stimulation of competition.[65] However, the reduction in price does not occur instantaneously and does not immediately achieve marginal cost. Rather, the reductions will be incremental based upon consumer reaction and the pricing of other vendors. The question then arises as to how vendors will adjust their prices to maximize profits.

If vendors were able to price products according to an idealistic game-theoretic model, then prices would decrease to marginal cost.[66] Imperfect information, however, forces vendors to choose from the myopically optimal method, the derivative-following method, or a combination thereof.[67] A seller adhering to the derivative-following method tests incremental increases and decreases in its price, continuing to adjust in the same direction until profitability decreases, and then reversing directions.[68]  The myopically-optimal seller fixes its price by using the available information about buyer characteristics and competitors' prices to maximize its profits.[69]  Simulations reveal that the derivative-following model leads to oligopolistic practices[70] and the myopically optimal model leads to cyclical price wars, with one vendor always offering the product at the marginal cost of production.[71]

---

[61] A "Nash Equilibrium," is a price vector from which a group of sellers maximizes their profits and have no incentive to deviate. However, the shopbot model presented of mixed consumers of type 1 and type 2 cannot yield a pure-strategy Nash Equilibrium, which could exist if all consumers were of type 2. At an equilibrium based on the shopbot model, at least one seller must offer *a* for $p<v$. However, this is not a stable condition and this seller has an incentive to deviate. Nevertheless, a mixed-strategy Nash Equilibrium can exist based on the shopbot model, which would lead to prices at the marginal cost of production if vendors could price according to a perfect game-theoretic model. *See id.* at *5-9.

[62] *See id.* at *5. Both elements of the demand function must also take into consideration whether *a* is priced at or below $v_b$.

[63] *See id.* at *5-6

[64] *See* Bakos, *supra* note 55, at 1690; Kephart, *supra* note 57, at *5.

[65] *See* Amy R. Greenwald & Jeffrey O. Kephart, *Shopbots and Pricebots*, 16 INT'L JOINT CONFERENCE ON ARTIFICIAL INTELLIGENCE 506 (1999); *infra* notes 33 & 61.

[66] *See* Bakos, *supra* note 55; Kuttner, *supra* note 56.

[67] *See* Greenwald & Kephart, *supra* note 65, at 510.

While an inevitable combination of these two pricing methods will most likely lead to price reduction, vendors will increase profits with quicker price setting.[72]  To accomplish quicker price resetting, vendors will invoke robots of their own to exploit metasites to learn other vendors' prices.[73]  However, given the vendor's incentive to make these requests as frequently as possible, vendor robot use will presumably overshadow consumer shopbot use.  The result of increased vendor robot use will be increased shopbot use, as metasites respond to vendor robot requests by deploying robots of their own.  This situation generates two undesirable results: 1) shopbots will begin to charge vendors for pricing information of other vendors and vendors in return will begin to charge shopbots for their own pricing information, essentially reinstating search costs, or 2) the presence of millions of pricing robots executing millions of requests may overtax the Internet to a significant degree.[74]

---

[68] *See id* at 510.

[69] *See id.* at *510*.  However, this strategy does not take into account competitor pricing strategy, but supposes that their prices will remain unresponsively fixed.  *See id* at 510.

[70] *See id.* at 510.  Even though derivative followers disregard seller and consumer characteristics, their behavior nevertheless results in a collusive state of oligopolistic pricing.  This occurs because downward pricing trends under this method prove to be far more fragile than upward trends.  "For example, if *A*'s price is currently above *B*'s but *A* reduces its price by an amount insufficient to undercut *B*, then *A*'s profits decrease, so that *A* raises its price in subsequent time steps."  *Id.* at 510.

[71] *See id.* at 509-10.  Under the myopically optimal pricing scheme, regardless of the initial distribution of the pricing vector, a configuration soon appears where all sellers price significantly above the marginal cost of production.  This state sparks one seller to reduce its price by a minimal amount.  This trend of reduction is repeated among the sellers until it would be unprofitable for the next seller to further reduce its price and instead it resets its price to the original monopolistic value.  The other sellers follow the same process, thereby generating a cyclical price war.

[72] *See id* at 5l0; *see generally* Kephart & Greenwald, *supra* note 57, at *20-32.

[73] *See* Greenwald & Kephart, *supra* note 65, at 511.

[74] *See id.* at 511.

Therefore, economic modeling of metasites suggests a cautious approach to the use of shopbots and the development of a robot pricing protocol to avoid harmful price wars.

### 2. Other Factors Causing Above Marginal Cost Prices

Even if reduced search costs encourage market efficiency, factors unrelated to search costs maintain price dispersion.[75] Empirical studies illustrate that price dispersion in Internet markets parallels price dispersion in traditional brick and mortar systems.[76] For example, Brynjolfsson and Smith found that prices for compact discs and books offered for sale through online vendors exemplify Internet price dispersion: prices differed as much as 50% for identical products and the average price dispersion was 29%.[77] Even if one maintains that lower search costs consequentially reduce prices, other factors manifest in Internet markets neutralize the effect of diminished search costs and contribute to conspicuous price dispersion.

Product heterogeneity constitutes an understandable source of price dispersion. It is logical that two distinct products generate two different prices. However, assuming that product heterogeneity only encompasses physical characteristics of products, it fails to explain the significant price dispersion in Internet markets. Many of the examined Internet vendors offered identical products for sale, such as the same book or same compact disc.[78] While these products may be physically identical, product heterogeneity also takes into account services, which are attached to the product.[79] For example, a favorable return policy may distinguish a book of one seller from the book of a seller with a less beneficial return policy. Nevertheless, these primary service characteristics tend not to vary among online retailers or are negatively related to price.[80] Therefore, price dispersion exists even in a market of homogeneous products.

---

[75] Price dispersion results when different prices exist for the same item at the same time. As noted above, the traditional Bertrand economics model assumes that as search costs are eliminated, price dispersion will vanish.

[76] *See* Joseph P. Bailey, Intermediation and Electronic Markets: Aggregation and Pricing in Internet Commerce (1998) (unpublished Ph.D. dissertation, Massachusetts Institute of Technology) (finding high price dispersion and higher costs generally for books, software and CDs); Erik Brynjolfsson & Michael Smith, *Frictionless Commerce? A Comparison of Internet and Conventional Retailers* 46 no. 4 MGMT SCIENCE *passim* (2000) (Finding lower prices but high dispersion for Internet vendors of books and CDs); Erick K. Clemons et al., *The Nature of Competition in Electronic Markets: An Empirical Investigation of Online Travel Agent Offerings*, *passim* (2000) (unpublished manuscript) (finding high price dispersion for online airplane ticket sales). When I performed my search for *Mary Poppins, see supra* note 41, the 34 results I received ranged in price from $11.20 to $28.47 (at Amazon.com, *see infra* Part II.C).

[77] Brynjolfsson & Smith, *supra* note 76 at *passim*.

[78] *See id.* at 27.

[79] *See* Smith et al., *supra* note 54, at *9.

Evidence suggests that consumer convenience also contributes to price dispersion among online vendors.[81] Time-sensitive consumers may value a seller's computer interface, which provides uncomplicated product location and evaluation tools.[82] Sellers that offer these aids may then charge consumers a premium for the convenience.[83] Factors relating to convenience among online retailers include facility of search mechanisms, product suggestion devices,[84] product reviews,[85] product samples,[86] and rapid checkout procedures.[87] These convenience elements are distinct from the physical characteristics (or accompanying services) encompassed within the product heterogeneity factor. Because these convenience tools are separate from the physical product, retailers should not gain a pricing advantage because a consumer could exploit the convenience offered by one vendor yet purchase from another. However, high search and switching costs may compel a consumer to remain with the vendor that offers the convenience tools. In addition to convenience tools, website design characteristics, such as background wallpaper, influence consumers' perception of a product.[88] Therefore, construction of an agreeable and efficient online purchasing forum may grant vendors a license to charge a premium over sellers that fail to offer competing convenience, leading to price dispersion unrelated to search costs.

---

[80] Brynjolfsson & Smith, *supra* note 76 at *29

[81] *See id.* at *10; DeLong & Froomkin *supra* note 43, at 26.

[82] *See* Alan E. Wiseman, *Economic Perspective on the Internet*, POLICY PAPERS *22 (Fed. Trade Comm'n July 31, 2000); DeLong & Froomkin, *supra* note 43, at 26, 29-32.

[83] *See* Wiseman, *supra* note 82, at *22.

[84] For example, Amazon.com employs several mechanisms to provide consumers with suggestions. A consumer can enter general criteria and Amazon.com will then offer selections based on these criteria. Furthermore, Amazon.com will suggest products based on past purchases of a specific consumer

[85] For example, Amazon.com displays both professional and consumer critique of its products, viewable adjacent to the basic product information.

[86] For example, Amazon.com offers consumers audio clips from tracks of musical selections offered for sale.

[87] Online retailers may offer "one-click" checkout, where a consumer's shipping and payment information is stored by the vendor to facilitate the administrative aspects of purchases. *See generally* Smith et al., *supra* note 54, at *10.

Branding and consumer-trust may also preserve price dispersion in the absence of search costs. The economic model described above[89] assumes that when a consumer exploits a metasite, he or she will then purchase from the vendor that offers the lowest price. Shopbot executives, however, have revealed that metasite visitors frequently purchase from recognized brands, even when these vendors present a substantially higher price than other retailers[90] In the nascent Internet market, trust and brand play an important role "because of the spatial and temporal separation between buyers and sellers imposed by the medium."[91] Because transactions for physical goods over the Internet involve disjointed payment and delivery, consumers may be willing to pay a premium to use a vendor that the consumer can trust with credit card information and that the consumer can rely upon for effective delivery. Furthermore, the general aversion to personal data mining associated with online retailers suggests that consumers will opt for a trusted brand, which they believe will safeguard their personal information.[92] Factors such as online communities located on the retailer's website, links to the retailer from other trusted websites or portals, extensive advertising and existing brick and mortar brand names contribute to Internet brand recognition and the ability to charge and receive higher prices.[93]

Vendors may also succeed in charging higher prices when they have implemented loyalty programs, which effectively lock in the consumer.[94] These programs offer consumers incentives to remain with the vendor and discourage use of a different retailer that may price lower. For example, a retailer may offer the equivalent of frequent flier miles, where the consumer accumulates bonuses with each successive purchase.

---

[88] *See* Naomi Mandel & Eric Johnson, *Constructing Preferences Online: Can Web Pages Change What You Want*? (unpublished paper), *available at http://fourps.wharton.upenn.edu/~naomi/construct.htm* (1998).

[89] *See supra* Part II.C.1.

[90] *See* O'Rourke *supra* note 7, at 1974. Well branded Internet vendors such as Amazon.com and CDNow charge an average of 12% higher than lesser known vendors, yet are frequently selected by consumers, even when the consumer is aware that the item is offered for less. *See id.*

[91] Smith et al., *supra* note 54, at *12.

[92] *See generally* Federal Trade Commission, *FTC announces Settlement with Bankrupt Website, Toysmart.com, Regarding Alleged Privacy Policy Violation*, (July 21, 2000), *available at* http://www.ftc.gov/opa/2000/07/toysmart2.htm (describing claims that a company misrepresented to consumers that their personal information would not be disclosed to third parties).

[93] *See* Smith et al., *supra* note 54, at *13.

[94] *See id.* at *13; DeLong & Froomkin, *supra* note 43, at 29-32.

Additionally, online vendors may utilize collaborative filtering tools, which compare a consumer's past selections with those of other consumers and generate recommendations based on these data.[95]  Given these loyalty programs, a consumer switches vendors at a cost equal to the value of the extras.  Furthermore, consumer lock-in may occur in the absence of direct loyalty programs, through the ease of use of the vendor's interface.  A consumer familiar with one online retailer's website may avoid switching to an unfamiliar website.  Therefore, lock-in mechanisms, both intentional and unintentional may contribute to price dispersion.

Finally, price discrimination provides another source of price dispersion unrelated to search costs.  Price discrimination exists when markets have been segmented between high and low valuation consumers, "such that the sellers can post a high price that high demand consumers will find attractive given their knowledge about the products in question."[96]  To illustrate price discrimination consider a product $a$, with a marginal cost of production of 9.  Consider 10 consumers Q, all of whom are willing to pay up to 13 for $a$.  Further, consider 1 consumer $R$, who is willing to pay up to 60 for $a$.  How does the vendor price $a$?  To maximize profits, the vendor will price $a$ at 60, sell it to $R$, and realize a profit of 51.  If the vendor were able to sell $a$ for 13 to $Q$ but at 60 to $R$, the vendor would realize a profit of (10*4)+(1*51)=91.  Thus, retailers may achieve greater income through price discrimination.

Two characteristics of the Internet make price discrimination more practicable than in brick and mortar markets.  First, through the Internet, vendors may easily track consumer characteristics.[97]  Second, lower menu costs allow vendors to change prices with greater ease.[98]  An example illustrates price discrimination on the Internet.  Books.com offered consumers a price matching policy.  It presented customers with a mechanism that compared its price for a specific item with the prices of several other vendors, and changed its price if another vendor offered the same item for a lower price.[99]  While initially this strategy appears to provide economic efficiency, it in fact price discriminates.  Books.com failed to maintain the new price for future transactions.  Thus, a consumer who elected not to engage in the price comparing process risked missing the pricing benefit, but avoided a potentially lengthy price comparison procedure and possible fruitless result.[100]  While price discrimination obviates economic efficiency, as it necessarily leads to different marginal rates of substitution, it may offer some consumers with lower prices and the ability to purchase where none existed before.

---

[95] *See* DeLong & Froomkin, *supra* note 43, at 32-35.

[96] Wiseman, *supra* note 82, at *22.

[97] *See id.* at *22 (citing Hal R. Varian, *Price Discrimination, in* HANDBOOK OF INDUSTRIAL ORGANIZATION (R. Schmalensee & R. D. Willig eds., 1989) (noting that a necessary condition for price discrimination to occur is the ability to sort customers)).

[98] *See id.* at 23; *see also* Greenwald & Kephart, *supra* note 65, at 511.

Therefore, even if the reduced search costs of the Internet would consequently lead to lower prices, other factors impede this reduction.  Product heterogeneity, convenience tools, trust in brands, consumer lock-ins, and price discrimination all may contribute to price dispersion on the Internet, unrelated to search costs.

<u>D. Summary of Economic Analysis</u>

The economic analysis of metasites yields surprising observations.  First, it is often in the sellers' interests to initially limit shopbot extraction of their pricing and product information, and sellers will presumably take steps to effect this preference.  Second, economic modeling demonstrates that shopbot use will not inevitably lead to lower prices and economic efficiency, but rather will likely induce price wars characterized by uncontrolled and damaging robot use.  Finally, in addition to search costs, other factors contribute directly to price dispersion, negating any reduction in price by diminished search costs.

Because comprehensive shopbot access would overly tax a system without reducing prices, and potentially redistribute income in a seller- and consumer-detrimental manner, a combination of legal and technical mechanisms that will safeguard the Internet markets in the transition to metasite use is warranted.  The economic modeling of metasites reveals that the pacing of price resetting is a crucial issue directly related to robot proliferation.  Given established sellers' reluctance to participate in price indexing, if they were able to invoke and rely upon legal and technical mechanisms to prohibit robot use on their websites, this would curtail the potential threat of system overload while permitting metasite functioning.[101] With the advancements of metasites to compare elements in addition to price, the other factors producing price dispersion will fade, assuming they are in fact specious, thus leading to lowered search costs, without robot proliferation.[102] Therefore, as metasites become a staple for consumers, branded retailers that had prevented shopbot access will find it in their interest to permit shopbot indexing.[103] In this fashion, we realize a complete transition to metasite use, while avoiding cyclical and taxing price wars, by establishing a time control mechanism.[104] While this end could be accomplished by legally requiring online retailers to permit robot access and then mandating and enforcing a shopbot protocol[105], as demonstrated, seller limitation of shopbots, through contract and technical standards, appears to offer the same result in an indirect manner.[106]

---

[99] *See* Wiseman, *supra* note 82, at *23.
[100]  A study of Books.com found that the average savings to a consumer who used the price comparison mechanism was $.15 and that the procedure could take up to 1 minute.  Thus, in essence, Books.com

## III. CONTRACT LAW AND ROBOT EXCLUSION

The explosive growth in personal computer and Internet use has ushered in a new breed of contracts. Gone are the handshake deals and ornate signatures of yore, replaced by shrinkwrap and clickwrap agreements. The hazy realm of intellectual property rights in computer information has motivated data owners to eschew governmental protection

---

discriminated on the basis of convenience and potential savings. It apparently found that enough consumers would find the process too time-consuming to justify the savings. *See id.* at *23.

[101] Presumably metasites would continue to index retailers for which the benefits of price indexing outweighed the negative aspects, specifically the less branded retailers.

[102] If these factors did not wane, but continued to cause price dispersion, then a static market would continue to exist, with consumers choosing well-branded but more expensive retailers, though the desired product is offered for less by other vendors. This presumably is a preferred state to the cyclical price wars, which would occur without this legal mechanism.

[103] When the elements other than search costs that affect pricing disappear, established retailers will find it necessary to compete with indexed retailers because consumers will otherwise simply select the lowest priced indexed vendor.

[104] Presumably, sellers who have avoided metasite price indexing will switch only when consumers are wholeheartedly relying on metasites and the vendors who are listed by the metasites offer products for substantially less than non-listed vendors' marginal cost of production. At this point, vendors will not need to engage in price resetting race, as the prices have already been set. The taxing price wars predicted in the economic model are thus avoided by temper the transition to metasite use.

[105] For example, the law could impose criminal or civil liability for one who failed to heed a robot exclusion header.

[106] Other remedies that retailer might take to protect profits would have more harmful effects on economic functioning. Sellers might charge user fees for access to the market, thus compensating for lost profits,

and embrace security through private agreement.[107]  Initially, to prevent the pirating of software, companies implemented shrinkwrap contracts, or unsigned licensing agreements accompanying the sale of computer software, for the purpose of binding the purchaser when he or she opened the package.[108]  Clickwrap agreements soon followed, referring to contractual terms present on websites.[109]  These web-based contracts range from chat-room access agreements to software-downloading arrangements and include terms ranging from indemnity to forum selection.[110]  While some websites induce users to express consent by having them click on an "I accept" button, other websites simply condition use of the site on an implicit acceptance of displayed contractual terms.[111]

If such clickwrap agreements are valid, a website could theoretically prevent robots from indexing the site's data by displaying a simple notice, which reads: "by accessing this website, the user agrees not to employ any robot to copy the content of the website."[112]  Can it be that simple?  Were all the arguments invoking trespass to chattels and the CFAA unnecessarily complicated?  Three elements may stand as obstacles to the enforceability of robot restriction contracts: 1) the doctrines of consent, contract of adhesion, and unconscionability; 2) copyright preemption; and 3) whether robots may assent to contract formation.  Despite these ostensible barriers, both contract case law and the Uniform Computer Information Transactions Act indicate that robot restriction contracts are enforceable.

---

raise the costs of acquiring the information, or increase product differentiation.  *See* Wiseman *supra* note 70, at 18; Bakos, *supra* note 55, at *23.

[107] *See* James Boyle, *Cruel, Mean or Lavish? Economic Analysis, Price Discrimination and Digital Intellectual Property*, 53 VAND. L. REV. 2007, 2010 (2000).

[108] *See* ProCD, Inc. v. Zeidenberg, 86 F.3d 1447, 1449 (7th Cir. 1996); Katy Hull, *The Overlooked Concern with the Uniform Computer Information Transactions Act*, 51 HASTINGS L. J. 1391, 1393 (2000).

[109] *See* Hotmail Corp. v. Van$ Money Pie, Inc. No. C 98-20064, 1998 WL 388389, at *5 (N.D. Cal. April 16, 1998); Hull, *supra* note 108, at 1393.

[110] *See* Dawn Davidson, *Click and Commit: What Terms Are Users Bound to When They Enter Web Sites?*, 26 WM. MITCHELL L. REV. 1171, 1180 (2000).

[111] *See* Hull, *supra* note 108, at 1393.  For example, users of eBay must agree to a seven page contract and click on an "I accept" button positioned at the end of the agreement.  See *Ebay*, 100 F. Supp. 2d 1058, 1060.  However, Register.com "imposes conditions on the access to and end use of data contained in its WHOIS database.  It publishes those terms of use on the home page of its Internet website and conditions entry into the database on assent to those terms."  *Register.com* 2000 WL 1855145, at *5.

[112] While, this presents the most extreme point of clickwrap enforcement, recent case law supports its validity.  Nonetheless, a more likely scenario is for websites to rely on a robot exclusion header to bind robots to the terms of website use.  By relying on the Robot Exclusion Standard, websites essentially force robots to face the terms of the contract before accessing the content of the website.  *See supra* note 122.

<u>A. Shinkwrap and Clickwrap Caselaw</u>

While for years, shrinkwrap licenses were of questionable legality, *ProCD v. Zeidenberg* marked a dramatic shift in the courts' thinking about these contracts. In *ProCD*, the defendant purchased a software compilation of telephone numbers, extracted the records from the database and incorporated them into his own computer program, thereby breaching the shrinkwrap agreement.[113] Judge Easterbrook, writing for the 7[th] Circuit, declared that manufacturers might enter into private contracts limiting the use of their products.[114]

Many argue that shrinkwrap and clickwrap agreements constitute unenforceable contracts of adhesion, where the offeree is frequently unaware he or she is entering into a contract.[115] The court in *ProCD* disregarded contract of adhesion and lack of consent arguments, noting that the offeror, as master of the offer, may invite acceptance by conduct.[116] Furthermore, the offeror may dictate the category of action or conduct that constitutes an acceptance.[117] Here, ProCD offered a contract that the consumer could choose to accept by using the software. Furthermore, Judge Easterbrook noted in a prior case that a contract "need not be read to be effective; people who accept take the risk that the unread terms may in retrospect prove unwelcome."[118] The court pushed contextual unconscionabiltiy concerns aside by stressing that "shrinkwrap licenses are enforceable

---

[113] *See* 86 F.3d at 1449-50.

[114] *Id.* at 1455.

[115] *See* Davidson, *supra* note 110, at 1194-1201; Mark A. Lemley, *Shrinkwraps in Cyberspace*, 35 JURIMETRICS J. 311, 317-18 (1995); J.H. Reichman & Jonathan A. Franklin, *Privately Legislated Intellectual Property Rights: Reconciling Freedom of Contract with Public Good Uses of Information* 875, 911 (1999). The humorist Dave Barry satirized the length and complexity of shrinkwrap and clickwrap agreements, commenting:

> By Breaking this seal, the user hereinafter agrees to abide by all the terms and conditions of the following agreement that nobody ever reads, as well as the Geneva Convention and the U.N. Charter and the Secret Membership Oath of the Benevolent Protective Order of the Elks, and such other terms and conditions, real and imaginary, as the Software Company shall deem necessary and appropriate, including the right to come to the user's home and examine the user's hard drive, as well as the user's underwear drawer if we feel like it, take it or leave it, until death do us part, one nation indivisible, by dawn's early light, in the name of the Father, the Son, and the Holy Ghost, finders keepers, losers weepers, thanks you've been a great crowd, and don't forget to tip your servers.

Garry L. Founds, *Shrinkwrap and Clickwrap Agreements: 2B or Note 2B?*, 52 FED. COMM. L.J. 99, 100 (1999) (quoting DAVE BARRY, DAVE BARRY IN CYBERSPACE 98 (1996)).

[116] *See ProCD* 86 F.3d at 1452.

[117] *See id.* at 1452.

[118] Hill v. Gateway, 105 F.3d 1147, 1148 (7th Cir. 1997).

unless their terms are objectionable on grounds applicable to contracts in general."[119]
Clickwrap agreements have received similar, but limited, judicial treatment. In *Hotmail Corp. v. Van$ Money Pie, Inc.*, the court found that Hotmail was likely to succeed on its breach of contract claim when the defendant failed to abide by Hotmail's online service agreement.[120] Furthermore, in *Caspi v. Microsoft Network*, the court affirmed the validity of a forum selection clause within an online agreement.[121] Finally, the court in *Register.com* enforced a clickwrap agreement where use of the plaintiff's website was conditioned upon assenting to terms posted on the site, requiring no affirmative manifestation of assent.[122]

Similar to the contract in *ProCD*, a robot restriction agreement would protect the unauthorized use of factual data, namely prices. Furthermore, a website could implement a consent scheme similar to that validated in *Register.com*, where use of website was conditioned upon consenting to terms of the contract. Despite *Register.com*'s holding, the automatic nature of data robots suggests a more prudent approach would be to incorporate the Robot Exclusion Standard within the clickwrap template. Under this scheme, a data robot would be forced to confront and process robot restriction terms. Therefore, given the direct parallels between robot restriction contracts and the shrinkwrap agreement in *ProCD*, and the supplemental weight of the clickwrap caselaw, robot restriction contracts appear enforceable under contemporary contract law.

Copyright preemption presents another potential obstacle to the enforcement of robot restriction contracts through state contract law. Copyright law attempts to maintain a balance between the creators' and users' interests by protecting works of authorship, but leaving elements such as ideas, facts and procedures within the public domain.[123] The question thus arises whether this balance is subject to variation by contract or whether the balance embodies unwavering federal policy positions. In other words, may

---

[119] *ProCD*, 86 F.3d at 1448.
[120] *See Hotmail* 1998 WL 388389, at *5.
[121] *See* Caspi v. Microsoft Network, 732 A.2d 528 (N.J. Super. Ct. App. Div. 1999).
[122] *See Register.com*, 2000 WL 1855145, at *7. While Verio argued that the contract was unenforceable because Verio was not asked to click on an icon indicating it accepted the terms, the court found that the presence of clearly posted contractual terms was sufficient. *See id.* at *7.
[123] *See* Founds, *supra* note 115, at 103-06; Dennis S. Karjala, *Federal Preemption of Shrinkwrap and On-line Licenses*, 22 DAYTON L. REV. 511, 514-21.

private parties create intellectual property rights through contract where no right existed before? In the context of robot restriction agreements, the issue becomes whether the factual pricing information, which is planted in the public domain by copyright law, may be rendered proprietary through contract.

In *ProCD,* the district court found that it would alter the delicate balance of copyright law to allow parties to contract around it.[124] Judge Easterbrook reversed, however, finding that the contract rights were not "equivalent to any of the exclusive rights within the general scope of copyright."[125] The court reasoned that while copyright law provides a right against the world, contract rights affect only the parties to the contract and that "strangers may do as they please."[126] The court further illustrated this distinction by emphasizing that if somebody found a copy of ProCD's database on the sidewalk, the licensing agreement would not limit the finder.[127] Therefore, Judge Easterbrook concluded that, "whether a particular license is generous or restrictive, a simple two-party contract is not equivalent to any of the exclusive rights within the general scope of copyright and therefore may be enforced."[128] For similar reasons, copyright law would presumably not preempt a robot restriction agreement under contract law.

Whether robots may enter into contracts and bind a third party, i.e., the robot's implementer, presents a final obstacle to the enforceability of robot restriction agreements. Some have turned to agency law principles to address this concern.[129] Agency describes a relationship between two parties where an agent party consents to act on behalf of and under control of the principal party.[130] When an agent acts according to his or her authority, the agent has the power to affect the legal status of the principal to the same extent as if the principal had acted.[131] As a result, when an agent enters into a contract with a third party, that third party may subsequently enforce the contract against the principal.[132] Therefore, under agency law, it appears that a robot that enters into a clickwrap agreement, either by clicking on an "I accept button," or disregarding the express protocol set forth in a robot exclusion header, binds the person who designed and implemented the robot.[133]

---

[124] *See* ProCD, Inc. v. Zeidenberg, 908 F. Supp. 640, 655 (W.D. Wis. 1996), *rev'd* 86 F.3d 1447 (7th Cir. 1996).
[125] *ProCD,* 86 F.3d at 1454.
[126] *Id.* at 1454.
[127] *Id.* at 1454.
[128] *Id.* at 1455. But Judge Easterbrook's stance on copyright preemption has generated significant controversy. For example, Dennis Karjala draws a distinction between widely accessible and generally inaccessible works. For widely accessible works, Professor Karjala argues that the copyright scheme does not merely serve as a default position that may be reordered through private contract. Rather, "to allow private reordering of these rights by agreement between the copyright owner and the direct users . . . would be to ignore the interest of the general public that the copyright balances have been so carefully designed to protect. Karjala, *supra* note 123, at 518-19. Furthermore, Judge Easterbrook's opinion has been criticized for failing to address whether the enforcement of state contract law may be preempted by the Supremacy Clause of the Constitution as opposed to Section 301 of the Copyright Act, which he focuses on. *See id.* at 522.

Therefore, the doctrines of assent and contract of adhesion, copyright preemption, and robot contracting fail to present concrete obstacles to the enforceability of robot restriction agreements under current contract common law.

### B. UCITA and Robot Restriction Agreements

Spurred by the need for uniformity within the realm of non-tangible information contracting, the National Conference of Commissioners on Uniform State Laws ("NCCUSL") composed and disseminated the Uniform Computer Information Transaction Act ("UCITA").[134] UCITA, a model contract law to be adopted by individual states, governs agreements that involve "computer information transactions,"[135] and has garnered significant support,[136] but also engendered strident opposition.[137] Computer information refers to "information in electronic form which is obtained from or through the use of a computer . . ."[138]  UCITA's coverage includes contracts to license software, create computer information, distribute computer information, and gain access to online databases.[139]  The concept of a robot restriction contract appears to fall squarely within UCITA's definition of a viable access agreement as "a contract to obtain by electronic means access to, or information from, an information processing system of another person, or the equivalent of such access."[140]  Furthermore, UCITA expressly addresses the three concerns present at common law: 1) the doctrines of consent and contract of adhesion, 2) copyright preemption, and 3) robot assent.

---

[129] *See* Middlebrook & Muller, *supra* note 3, at *passim.*

[130] RESTATEMENT (SECOND) OF AGENCY § 1 (1958).

[131] *See* id. at § 12; Middlebrook & Muller, *supra* note 3, at 356.

[132] *See* RESTATEMENT (SECOND) OF AGENCY at § 140; Middlebrook & Muller, *supra* note 3, at 356.

[133] Furthermore, the recently enacted federal E-SIGN statute states that "a contract . . . may not be denied legal effect, validity or enforceability solely because its formation, creation or delivery involved the action of one or more electronic agents so long as the action of any such electronic agent is legally attributable to the person to be bound." 15 U.S.C. 7001(h). While E-SIGN does not expressly validate contracts formed by robots, it appears to yield to their enforceability.

[134] *See* Carlyle C. Ring, Jr., *Uniform Rules for Internet Information Transactions: An Overview of Proposed UCITA*, 38 DUQ. L. REV. 319, 321-23 (2000).

[135] *See* UNIFORM COMPUTER INFORMATION TRANSACTIONS ACT (2000) [hereinafter UCITA].

[136] *See, e.g., generally* Mary Jo Howard Dively, *A Survey of the Electronic Contracting Rules in the Uniform Electronic Transactions Act and the Uniform Computer Information Transactions Act*, 38 DUQ. L. REV., 209 (2000); Garry L. Founds, *supra* note 115.

[137] *See, e.g., generally* Hull, *supra* note 108; Apik Minassian, *The Death of Copyright: Enforceability of Shrinkwrap Licensing Agreements* 45 UCLA L. REV. 569 (1997).

UCITA offers liberal rules of contract formation, providing that a contract may be effected by any method sufficient to show agreement, including offer and acceptance or conduct by the parties.[141]  "An offer to make a contract invites acceptance in any manner and by any medium reasonable under the circumstances."[142]  The question then becomes what constitutes a reasonable acceptance of robot restriction terms.  While the affirmative clicking on an "I agree" button would presumably satisfy this provision of UCITA, it is questionable whether the conditioning of website use on assent to posted terms would prove reasonable under the circumstances.  Rather, the use of a robot exclusion header, in combination with posted terms would prove more efficient and reliable.  Use of the Robot Exclusion Standard would compel robots to confront contractual terms without beleaguering legitimate website users with burdensome consent provisions.

UCITA explicitly asserts that federal preemption applies.[143]  As with all state laws, those provisions of UCITA that are preempted by federal law become invalid to the degree of the preemption.[144]  Nonetheless, UCITA renders shrinkwrap and clickwrap agreements generally enforceable, despite the inherent nature of information licensing to implicate values embodied in federal copyright law.[145] The primary goal of clickwrap and shrinkwrap agreements is to limit the copying and use of proprietary material.[146] Therefore, while UCITA claims to be neutral in the debate over whether copyright law should preempt private information contracts,[147] it in fact noticeably slants toward Judge Easterbrook's view in *ProCD*. Raymond Nimmer, the reporter for UCITA, claims that copyright law functions as a default standard, which may be contracted around by consenting parties. Therefore, if UCITA's implicit endorsement of clickwrap agreements that implicate copyright privileges becomes established, robot restriction agreements would not be stymied by federal preemption under UCITA.

---

[138] UCITA § 102.

[139] See Pratik A. Shah, *Berkeley Law Journal Annual Review of Law and Technology: The Uniform Computer Information Transactions Act,* 15 BERKELEY TECH. L.J. 85, 89 (2000).

[140] UCITA § 102.

[141] UCITA § 202; *see also* Founds, *supra* note 115, at 102.  While UCITA contains a special provision governing mass-market licenses, robot restrictions would not constitute such agreements under § 209.  And even if one could argue that robot restriction agreements represent mass-market licenses, the extra safeguards provided would prove inconsequential in this context.

[142] UCITA § 203.

[143] UCITA § 105(a).

[144] BLACK'S LAW DICTIONARY 1177 (6th ed. 1990)

Finally, UCITA strengthens the position of the ability of data robots to form contracts. UCITA confirms that robots or electronic agents may enter into contracts with other electronic agents or individuals,[148] and furthermore contains an attribution rule, which states that an electronic authentication is attributed to the person who implemented the electronic agent.[149] "A person that uses an electronic agent that [he or she] has selected for making an authentication, performance, or agreement, including manifestation of assent, is bound by the operations of the electronic agent, even if no individual was aware of or reviewed the agent's operations or the results of the operations."[150] Therefore, if a pricebot validly forms a contract pursuant to UCITA, its implementer is bound by the contract's terms.

An electronic agent manifests assent if it authenticates a contract or engages in operations or conduct that indicate acceptance of the contractual terms.[151] Additionally, proof that a robot obtained or used information where the robot must have engaged in specific conduct to access the information, verifies assent to a contract that limits access to the information.[152]  Moreover, a website provides an electronic agent with the opportunity to review a contract if it makes it available in a manner that a reasonably configured electronic agent would react to.[153]  Therefore, placing the contractual terms of a robot restriction agreement within the robot exclusion header would notify a reasonably configured robot of the website's policy. [154]  Consequently, UCITA's strong robot attribution rules and its robot assent provisions appear to sanction pricebot restriction agreements when implemented through a robot exclusion header.

---

[145] *See supra*; UCITA § 209. As illustrated, UCITA establishes a template for mass-market contracts involving computer. This in essence is the definition of a shrinkwrap or clickwrap agreement.
[146] *See* Shah, *supra,* note 139, at 97. Shrinkwrap and clickwrap agreements developed to protect the unlicensed copying and distribution of software, as well as a mechanism to circumvent the first sale doctrine have traditionally involved intellectual property rights. As the Internet gains popularity as a retail medium, clickwrap agreements have progressed to include non-intellectual property aspects, such as a user agreement dictating appropriate conduct in a chat-room. *See* Hull, *supra* note 108, at 1393-94. Nevertheless, UCITA's mass-market licensing scheme was designed with the traditional shrinkwrap and clickwrap contracts in mind. *See* Founds, *supra* note 115, at 103.
[147] *See* Ring, *supra* note 134, at 326.
[148] UCITA § 206.
[149] UCITA § 213; *see also* Middlebrook & Muller, *supra* note 3, at 352.
[150] UCITA § 107.
[151] UCITA § 112.

Therefore, UCITA essentially sanctions the use of robot restriction agreements.  It provides liberal contract formation policies, implicitly approves of private contracting around federal copyright law, and offers liberal robot assent specifications.  A properly formulated robot restriction agreement, designed to prevent shopbots from gathering pricing information, thus appears to fall squarely within the realm of UCITA.

C. Summary and Conclusion of Current Law Regarding Robot Exclusion Contracts

Dicta in *Ebay* and *Register.com* suggest that web-based contracts, which users would encounter upon entering a website, would prove effective in preventing price data collection by shopbots.[155]  While elements of clickwrap agreements have sparked controversy surrounding these contracts' enforceability, contract common law and the momentum gained by UCITA promote the enforceability of properly couched robot restriction contracts.  Practically, a robot restriction agreement could either compel consumers to click on an "I accept" button or incorporate its terms through the code of a robot exclusion header.  By relying on the Robot Exclusion Standard, a website offers a legally viable shopbot restriction agreement without burdening legitimate human web users with unavoidable clickwrap language

IV. INTEGRATION OF PUBLIC POLICY, ECONOMIC EFFICIENCY, AND
CONTRACT LAW: THE NEED FOR A FAIR USE STANDARD

---

[152] UCITA § 112(d).
[153] UCITA § 112(d); *see also* Middlebrook & Muller, *supra* note 3, at 352.
[154] *See supra* Part I (describing robot exclusion headers and text files).
[155] *See Ebay* 100 F. Supp. 2d at 1060; *Register.com*, 2000 WL 1855145, at *5-8.

The combination of computing technologies and the Internet has revolutionized the production and dissemination of information, while obscuring the border between the public and private domains. The improvements in the calculation and compilation processes of factual information, and the drastic lowering of search costs, have facilitated access to all information.[156] Intellectual property law has traditionally recognized a dual function of information: on the one hand information should exist as a proprietary entity under the control of the creator, and on the other hand, information should function as an unprotected element to promote knowledge exploration and growth.[157] While federal intellectual property law establishes a balance between these two ends of the spectrum, the question arises whether the use of contract law as a proxy for intellectual property law disregards the public use aspect of information. In the context of robot restriction agreements, the question becomes whether denying robots access to pricing and product information frustrates valuable public use.

Contract common law and UCITA both support the enforceability of robot restriction agreements. This contracting regime thus places pricing and product information at the extreme proprietary end of the dual information spectrum; these legal systems permit uncomplicated formation of contracts that grant wide-ranging control of factual information and restrict even potentially beneficial public use of the data. The question then arises whether any beneficial use of product and pricing information outweighs the need to afford proprietary rights to the records. Economic analysis has revealed: 1) that unencumbered metasite activity leads to price wars, which may well result in mass system strain or price premiums; and 2) even if reduced search costs necessarily lead to lower prices, factors unrelated to search costs will maintain price dispersion. Therefore, economic analysis provides an incentive to lean in favor of proprietary interests over the public use of pricing and product data.

Nevertheless, it is not difficult to imagine beneficial uses of pricing and product data. First, while pricing data itself may prove of little use in the progress of education, science and research, data robots and metasites are far from limited to pricing information. Rather, data robots can potentially collect and compile information regarding any subject matter, including addresses and contact information, research results, and event and transportation scheduling. Furthermore, metasites continue to offer consumers increasingly more information in addition to price, such as warranty provisions, shipping processes, and consumer satisfaction. Thus, the ease of formation of robot restriction agreements may limit valid and worthwhile public access to an unlimited range of data. For example, research on factors unrelated to search costs, which maintain price dispersion, may promote the elimination of these factors (if they are in fact false), thereby allowing lower-search costs to lead to marginal cost of production pricing through controlled shopbot implementation.

---

[156] *See* Reichman & Franklin, *supra* note 115, at 884.

Second, the detrimental first prong of the economic analysis was predicated upon shopbots and metasites acting as profit-seeking enterprises. Under this model, metasites had incentives to satisfy as many queries as rapidly as possible, thus furthering disadvantageous price wars. However, a public interest, non-profit metasite might sidestep this obstacle by limiting its queries, thereby removing an essential link in the perpetration of retailer price-query/price-resetting functions.[158] Furthermore, if such a non-profit, public interest metasite existed, it might provide improved opportunities for the price-affecting factors unrelated to search costs to wane. Therefore, concrete public-use interests exist for product and pricing information.

How then should contract law permit acceptable public use of pricing and product information, while permitting retailers to control access to pricing data and thereby preserve the integrity of online markets? A potential solution would be the development of a technical standard modeled after the principles of the fair use defense to federal copyright infringement. The fair use defense permits courts to avoid rigid application of the law when it would unfairly restrict dissemination of useful works to the public.[159] Section 107 of the Copyright Act lists four factors to consider when deciding whether a

---

[157] *See id.* at 884-85.

[158] If for example a metasite updated its prices only once an hour or once a day, this would eliminate the possibility of retailers relying on metasites to check other retailer's prices and adjust their own prices accordingly.

Enforceability of robot restriction agreements, which rely upon the Robot Exclusion Standard would, as discussed, also result in a limitation of shopbot queries, thereby accomplishing the same result public use is justified: 1) the purpose and character of the use, 2) the nature of the protected work, 3) the effect of the use upon the potential markets for the work, and 4) the amount of the work used.[160]  A viable fair-use standard of robot accommodation could directly parallel the federal copyright provisions and include a condition limiting the frequency of robot use.  This technical standard could be incorporated within the robot exclusion header, which would already be used to couch the robot restrictive contractual terms.

Therefore, if one were to contemplate a non-profit metasite, which performed only a limited number of inquiries (as opposed to a commercial metasite that had an economic incentive to maximize its inquiries), such a use would satisfy a technical standard that focused on the character of the information collected and the process used to collect it.  The purpose and character of the use reflect strong economic policy.  The nature of the work, i.e., factual pricing data, engenders less sympathy if taken than other forms of protected work, because the substance of prices is inherently intended to be disseminated to consumers.  Furthermore, the technical standard could indicate the maximum frequency of robot queries that the fair use standard will tolerate.[161]  Therefore, a technical standard accommodating fair use shopbot operation would prove viable and beneficial in restoring the balance of the dual nature of information, while preserving the economic and proprietary incentives.

## V. CONCLUSION

While the Internet offers the potential to radically transform economic markets and consumerism, we must not be rash in assuming that by its nature it will inherently accomplish this goal.  Rather, the law should recognize that the reduced search costs of the Internet create as many dangers as they offer remedies.  While reduced search costs are inherently beneficial, to superimpose them upon a market and technological structure unprepared to accommodate them, while ignoring the potentially adverse consequences, could be as harmful as inhibiting the reduction in search costs in general. The uncertain direction of Internet markets encourages a more balanced acquiescence to pricebots and metasites. Enforceability of robot restriction contracts and the addition of a fair use technical standard offers a stable and evenhanded approach to encouraging economic efficiency and avoiding the dangers of reckless reliance on the Internet.

---

[159] *See* 17 U.S.C. § 107; Campbell v. Acuff-Rose Music, Inc. 510 U.S. 569, 577 (1994).
[160] *See* 17 U.S.C. § 107.

---

[161] The hypothetical non-profit metasite, because it moderates its robot use, could satisfy the robot frequency element within the technical standard. Contrarily, a commercial metasite will have economic incentives that urge it to deploy robots as frequently as possible.